\documentclass[10pt,preprint]{emulateapj}

\newcommand\brittion[2]{\mbox{#1\hspace{0.2em}{\rmfamily{#2}}}}

\shorttitle{Large-Scale Star Formation-Driven Outflows at $1<z<2$}
\shortauthors{Lundgren et al.}

\begin{document}


\title{Large-Scale Star Formation-Driven Outflows at $1<z<2$ in the 3D-HST Survey}


\author{Britt F. Lundgren\altaffilmark{1}, Gabriel Brammer\altaffilmark{2}, Pieter van Dokkum\altaffilmark{1}, Rachel Bezanson\altaffilmark{1}, Marijn Franx\altaffilmark{3}, Mattia Fumagalli\altaffilmark{3}, Ivelina Momcheva\altaffilmark{1}, Erica Nelson\altaffilmark{1}, Rosalind E. Skelton\altaffilmark{1}, David Wake\altaffilmark{1}, Katherine Whitaker\altaffilmark{1}, Elizabete da Cunha\altaffilmark{4}, Dawn K. Erb\altaffilmark{5},  Xiaohui Fan\altaffilmark{6}, Mariska Kriek\altaffilmark{7}, Ivo Labb{\'e}\altaffilmark{3}, Danilo Marchesini\altaffilmark{8}, Shannon Patel\altaffilmark{3}, Hans Walter Rix\altaffilmark{4}, Kasper Schmidt\altaffilmark{4}, Arjen van der Wel\altaffilmark{4}}





\altaffiltext{1}{Astronomy Department, Yale University, New Haven, CT 06511}
\altaffiltext{2}{European Southern Observatory, Alonso de C{\'o}rdova 3107, Casilla
19001, Vitacura, Santiago, Chile}
\altaffiltext{3}{Leiden Observatory, Leiden University, Leiden, The Netherlands}
\altaffiltext{4}{Max Planck Institute for Astronomy (MPIA), K{\"o}nigstuhl 17, 69117,
Heidelberg, Germany}
\altaffiltext{5}{Department of Physics, University of Wisconsin-Milwaukee, P.O. Box
413, Milwaukee, WI 53201, USA}
\altaffiltext{6}{Steward Observatory, University of Arizona, Tucson, AZ 85721, USA}
\altaffiltext{7}{Department of Astronomy, University of California, Berkeley, CA 94720, USA}
\altaffiltext{8}{Physics and Astronomy Department, Tufts University, Medford, MA, 02155, USA}




\begin{abstract}
We present evidence of large-scale outflows from three low-mass (log(M$_{*}$/M$_{\odot})\sim9.75$) star-forming (SFR $>4$ M$_{\odot}$ yr$^{-1}$)  galaxies observed at $z=1.24$, $z=1.35$ and $z=1.75$ in the 3D-HST Survey.  Each of these galaxies is located within a projected physical distance of 60 kpc around the sight line to the quasar SDSS J123622.93+621526.6, which exhibits well-separated strong (W$_{r}^{\lambda2796}\gtrsim0.8$\AA) \brittion{Mg}{II} absorption systems matching precisely to the redshifts of the three galaxies.   We derive the star formation surface densities from the H$\alpha$ emission in the WFC3 G141 grism observations for the galaxies and find that in each case the star formation surface density well-exceeds 0.1 M$_{\odot}$ yr$^{-1}$ kpc$^{-2}$, the typical threshold for starburst galaxies in the local Universe.  From a small but complete parallel census of the $0.65<z<2.6$ galaxies with $H_{140}\lesssim24$ proximate to the quasar sight line,  we detect \brittion{Mg}{II} absorption associated with galaxies extending to physical distances of 130 kpc.  We determine that the W$_{r}>0.8$\AA~ \brittion{Mg}{II} covering fraction of star-forming galaxies at $1<z<2$ may be as large as unity on scales extending to at least 60 kpc, providing early constraints on the typical extent of starburst-driven winds around galaxies at this redshift.  Our observations additionally suggest that the azimuthal distribution of W$_{r}>0.4$\AA~\brittion{Mg}{II} absorbing gas around star-forming galaxies may evolve from $z\sim2$ to the present, consistent with recent observations of an increasing collimation of star formation-driven outflows with time from $z\sim3$.
\end{abstract}


\keywords{galaxy evolution: general; quasar absorption lines: general}



\section{Introduction}

Outflowing galactic winds have been widely observed in galaxies from z$\sim$6 to the present and appear to play a fundamental role in galaxy evolution, regulating galactic star formation \citep[e.g.,][]{Sanders88, DiMatteo05, Hopkins05} and enriching the intergalactic medium (IGM) at high redshift \citep[e.g.,][]{Madau01, Scannapieco02}.  However, the enclosed gas mass, physical extent, and physical conditions required to trigger galaxy-scale outflows remain to be better quantified before a complete understanding of the contribution of winds to observed co-evolution of galaxies and the IGM may be achieved. 

Bright, background quasars provide one of the most effective tools for probing the cold gas content of galaxies and their extended halos \citep[e.g.,][]{BS69}, and metal absorption features in quasar spectra have been shown to probe enriched gas in a wide range of galaxy environments, extending to $\sim150$ kpc around foreground galaxies.   The low-ionization \brittion{Mg}{II} doublet ($\lambda\lambda$2796,2803\AA), a tracer of T $\sim10^{4}$ K photo-ionized gas, is particularly prolific, and is observable in the range $0.35<z<2.2$ in the optical and across six decades of neutral hydrogen column density \citep[e.g.,][]{Churchill2000}.  Thus, \brittion{Mg}{II} provides a common and sensitive probe of the distribution of enriched gas in galaxy halos, capable of tracing the most fundamental disk-halo processes: star formation-driven outflows and cold-mode accretion.

Despite the fact that tens of thousands of \brittion{Mg}{II} absorption line systems have been extracted from spectroscopic quasar surveys to date \citep[e.g.,][]{Prochter06, Quider11, York12}, our understanding of the origins and environments of these absorbers remains underdeveloped.  This has largely been due to the fact that at the redshifts where \brittion{Mg}{II} is most easily detected ($z>0.4$), the individual galactic hosts tend to be too faint or too close to the quasar to be properly resolved with ground-based observations.

Theoretically \brittion{Mg}{II} absorption is capable of tracing the interstellar medium of the host galaxy \citep[e.g.,][]{PW97}, star formation-driven outflows \citep[e.g.,][]{Norman96,Nulsen98}, and cold mode accretion from gas in the extended halos of galaxies \citep[e.g.,][]{Kacprzak10, Stewart11}. The currently favored paradigm implies that the absorber rest-frame \brittion{Mg}{II} equivalent width, W$_{r}$ (2796\AA), can be used to discriminate between these origins, with the highest equivalent widths being associated with disks and star formation-driven winds, and the lowest W$_{r}$ systems indicating more extended virialized gas. 

At intermediate redshifts, the observed anti-correlation between clustering strength (halo mass) and W$_{r}$ argues against the interpretation that the Mg II absorption systems are virialized  \citep{B06, L09, Gauthier09}; instead, the observed correlation between W$_{r}$  and star-formation rate (SFR) from stacking analyses implies that these systems can be explained by outflowing winds \citep{Zibetti07, NSM10, Menard11}.  This scenario is also consistent with observations of blue-shifted \brittion{Mg}{II} absorption features in the spectra of star-forming galaxies at similar redshifts \citep{Weiner09, Rubin10, Erb12, Martin12, Kornei12}. However, the aforementioned statistical studies of \brittion{Mg}{II} quasar absorption lines are not based on the direct detection of \brittion{Mg}{II} absorber host galaxies, and it has been suggested that both the stacking and clustering results could have other explanations \citep{TC08, Chen10a}.  Still, the detection of a strong azimuthal dependence of high-W$_{r}$ \brittion{Mg}{II} absorption within 50 kpc of disk-dominated galaxies at $z<1$ has added to the support for an outflow origin \citep{Bordoloi11, Kacprzak12, B12b}, given the similar geometry of bipolar outflows observed in local starbursts \citep[e.g.,][]{Heckman90, LH96, Cecil01, SH09}.

A number of studies have used deep imaging or spectroscopy to try to either directly or statistically identify \brittion{Mg}{II} host galaxies at $z>0.5$ \citep[e.g.,][]{LeBrun93, Steidel97, Nestor07, B07, Straka10, Chun10, Nestor11} resulting in a few hundred individual detections. While \brittion{Mg}{II} has been found in association with a wide range of galaxy types, the ultra-strong (W$_{r}>$2\AA~) absorbers seem to predominantly trace galaxies with high SFRs, at least at high redshift.

However, many galaxy hosts of ultra-strong \brittion{Mg}{II} remain undetected in even the deepest ground-based data. In the most sensitive survey to date, \citet{B07} detected just 66\% of the host galaxies of ultra-strong \brittion{Mg}{II} absorbers at $z\sim1$ using ultra-deep SINFONI IFU observations, and this number dropped to 20\% at $z\sim2$ \citep{B12a}. These results could indicate that either the \brittion{Mg}{II} host galaxies have SFRs below the detection limit, are beyond the SINFONI field of view (40 kpc at $z=1$), or lie too close to the quasar to be detected at ground-based resolution.  

In the low redshift universe ($z\lesssim0.5$), studies of directly detected \brittion{Mg}{II} absorber host galaxies have shown that the origins of \brittion{Mg}{II} are even less clear in the case of the more common population of absorbers with W$_{r}\la$2\AA.  Analyses of galaxies around detected \brittion{Mg}{II} absorbers \citep{Kacprzak11b} and of the absorption properties of galaxies close to quasar sight lines \citep{Chen10a, Chen10b} each find an inverse relation between W$_{r}$ and impact parameter at $z\lesssim0.4$ with a large ($\sim1$ dex) scatter, which can be reduced by accounting for host galaxy inclination and luminosity.  The SFRs of the host galaxies in each study provide little support for models in which winds driven by star formation in the host galaxy produce \brittion{Mg}{II} absorption with W$_{r}\la1$\AA, suggesting that the bulk of \brittion{Mg}{II} absorbers probes infalling gas from disk-halo processes in normal galaxies.  

While much of the evidence described above supports a picture in which high equivalent width metal absorbers trace star-forming galaxies at small angular separations from quasar sight lines, the high W$_{r}$ systems have typically been studied at high redshift where only the brightest most rigorously star-forming systems are likely to be identified; whereas the local universe studies, which probe galaxies to fainter limits, are statistically restricted to lower W$_{r}$ systems as well as galaxies with lower average specific SFRs. This potential bias combined with the uncertainties outlined above suggest that the interpretation of a correlation between W$_{r}$ and host SFR requires a deeper examination.

In an effort to begin resolving this problem, we have harnessed the peerless sensitivity and resolving power of the WFC3 G141 grism to study the absorption properties of a complete sample of galaxies surrounding the only known bright $z>2$ quasar within the 3D-HST survey \citep{PvD11,Brammer12}.  These data enable us to study, for the first time, the properties (redshifts, morphologies, azimuthal angles, SFRs, and stellar masses) of a small but complete sample of Mg II-selected galaxies at $z>1$.  Furthermore, we are also able to study the W$_{r}>0.2$\AA~absorption properties for a volume-limited sample of galaxies in the foreground of the quasar at virtually unlimited impact parameters (5 $\lesssim \rho$ [kpc] $\lesssim$ 450) and to extremely low SFRs (2 M$_{\odot}$ yr$^{-1}$), thus eliminating the most prohibitive biases of previous high redshift studies of the cold gas content of normal galaxies.

The observations contributing to this work are described in Section 2, and details of our analysis are given in Section 3.  We present a discussion of our results in Section 4, along with a discussion of the implications of these results, with regard to the evolving distribution of \brittion{Mg}{II} around galaxies from $z\sim2$.   Throughout this paper, we assume a flat $\Lambda$--dominated CDM cosmology with $\Omega_m=0.3$, $H_0=70$ km s$^{-1} $Mpc$^{-1}$, and $\sigma_8=0.8$ unless otherwise stated. 

\begin{table*}[t!]
\centering
\begin{minipage}{0.95\textwidth}
\begin{center}
\caption{Ancillary Deep Broad-Band Photometry\label{tbl-bbphot}}
\begin{tabular}{rrrrr}
\tableline\tableline
Filter & Telescope & Instrument & Observations & Reference  \\
\tableline\tableline
\hline
U & KPNO 4m & MOSAIC & 9-13 March 2002 & \citet{Capak04} \\
G & Keck I & LRIS B & 3 April 2003 & \citet{Steidel03} \\
Rs & Keck I & LRIS R &  3 April 2003 & \citet{Steidel03} \\
B, V, I, Z & {\it HST} & ACS &  {\it HST} Cycle 12 9583 & \citet{Giavalisco04} \\
J, H, Ks & Subaru & MOIRCS &   2006-2008 & \citet{Kajisawa11} \\
F140W & {\it HST} & WFC3 &  {\it HST} GO-11600 & Weiner et al. (in prep.) \\
3.6, 4.5, 5.8, 8.0$\mu$m & {\it Spitzer} & IRAC &  2004 & \citet{Dickinson03} \\

\tableline	
\end{tabular}
\end{center}
\end{minipage}
\end{table*}

\section{Observations}

\subsection{Galaxy Observations in the 3D-HST Survey}

The 3D-HST Survey \citep{PvD11, Brammer12} is a 600 arcmin$^{2}$ survey using the Hubble Space Telescope to obtain complete, unbiased low-resolution spectra for $\sim7,000$ galaxies at $1<z<3.5$ (Cycles 18 and 19, PI van Dokkum).  The 3D-HST observing strategy employs WFC3 G141 grism observations, paired with WFC3 F140W direct imaging in order to extract a nearly complete census of 2-dimensional spectra for objects in the survey area to a 5$\sigma$ limiting depth of H$_{140}\sim26.1$, comparable to the deepest ground-based surveys.  Data from the HST program GO-11600 (PI Weiner), which covers GOODS-N with the same depth in G141 and F140W, are also included as part of the 3D-HST survey.  It is the data obtained by this GOODS-N program that specifically contain the quasar sight line examined in this work.

The WFC3 G141 provides slitless spectroscopy over the wavelength range 1.10$-$1.65$\mu$m with a first-order dispersion of 46.5 \AA/pixel (R $\sim130$) and a spatial resolution of $\sim0.\arcsec13$, sampled with 0.\arcsec06 pixels.   These specifications enable the detection of H$\alpha$ emission in the redshift range $0.7<z<1.5$, [\brittion{O}{III}]$\lambda$5007 in the range $1.2<z<2.3$, and [\brittion{O}{II}]$\lambda$3727 for $2.0<z<3.4$.  The grism data has been reduced and analyzed using the aXe \citep{Kummel09} package for the extraction of slitless spectroscopy.  The reduction relies on first detecting sources from the F140W imaging, using SExtractor \citep{BA96}.  Further care is taken in the extraction to reduce contamination from nearby sources and sky background.  Details of this procedure are provided in \citet{Brammer12}.

In addition to the WFC3 imaging and grism data, the 3D-HST Survey employs ancillary broad-band photometry in the overlapping fields to determine the broader spectral energy distribution (SED) of each object detected in the WFC3 F140W imaging.  In the case of GOODS-N, the ancillary deep broad-band imaging ranges from the U-band through 8$\mu$m, as detailed in Table ~\ref{tbl-bbphot}.  The aperture photometry, to be detailed in \citet{Skelton12}, was performed after matching all images to the point-spread function of the F140W image, using an identical technique as \citet{L12a}.  The photometric data provides a means for scaling and tilting the 1D spectra, which optimizes the determination of the best-fit redshift and stellar population of each source.  The method of fitting the redshifts using the EAZY code \citep{Brammer08} are described in detail in \citet{Brammer12}.  

\begin{figure}[t!]
\epsscale{1.15}
\plotone{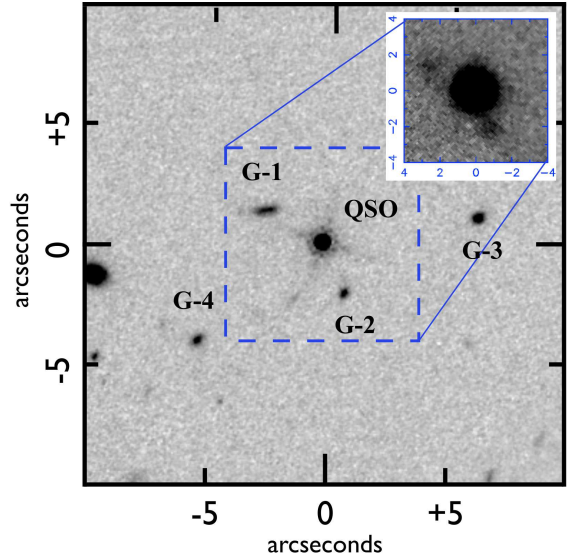}
\caption{A 20\arcsec x20\arcsec~cutout of a WFC3 F140W image from the GOODS-N grism program GO11600 (PI Weiner), centered on the m$_{r}$=20.4 quasar SDSS J123622.93+621526.6 with z=2.59.  Each of the four galaxies identified within 60 kpc of the quasar sight line are labeled G1$-$4.  Inset is a 8\arcsec x 8\arcsec~DEIMOS R-band image from the Team Keck Treasury Redshift Survey \citep[TKRS;][]{Wirth04}, illustrating the difficulty of using ground-based observations to obtain redshifts of faint galaxies with small angular separations around even this reasonably faint background quasar.  \label{fig:fov}}
\end{figure}

\subsection{Quasar Observations in the SDSS and TKRS}

Due to the relatively small area covered by the 3D-HST  Survey and the general sparseness of bright quasars, a search for quasars in the Sloan Digital Sky Survey \citep[SDSS;][]{York2000} DR7 quasar catalog \citep{Schneider10} within the 3D-HST footprint produced only one object: quasar SDSS J123622.93+621526.6, with $z=2.5946$ and m$_{r}=20.4$.  A 20\arcsec x20\arcsec~WFC3 F140W image from program GO-11600 of the field around this quasar is presented in Figure~\ref{fig:fov}.  The exceptional gain in resolution and depth of the WFC3 imaging in comparison to high-resolution ground-based imaging is illustrated by the 8\arcsec x 8\arcsec~R-band image from Keck II \citep{Wirth04}, shown as an inset in the figure. 

The signal-to-noise ratio (SNR) per resolution element of the SDSS DR7 spectrum of SDSS J123622.93+621526.6 is $\sim3$, which at the typical SDSS resolution of R $=1800$ \citep{Uomoto99} is sufficient only for detecting absorption features with W$_{r}\gtrsim1.5$\AA~at 3$\sigma$ significance.  As the \brittion{Mg}{II} equivalent width distribution function (dN/dW) in the SDSS is steep \citep{Nestor05}, most \brittion{Mg}{II} absorbers have W$_{r}<1$\AA.  We therefore required a greater depth than the limiting equivalent width of the SDSS observation in order to undertake a more comprehensive analysis of the foreground absorption in this particular quasar spectrum.

A spectrum of the same quasar with higher signal-to-noise is publicly available as part of the Team Keck Treasury Redshift Survey \citep[TKRS;][]{Wirth04}, a deep imaging and spectroscopic survey in the GOODS-N field using the Deep Imaging Multi-Object Spectrograph \citep[DEIMOS;][]{Faber03} on the Keck II telescope.  The TKRS contains spectra of 1440 galaxies and AGN, with wavelength coverage from 4600$-$9800\AA~and a FWHM resolution of $\Gamma\approx3.5$\AA.  Each slit mask had an on-source integration time of 3600s, resulting in a mean per pixel SNR of 11.7 for the quasar examined in this work (TKRS GOODS-N ID: J123622.94+621527.0).  Though the DEIMOS resolution is slightly worse than that of the SDSS, the higher SNR enabled detection of absorption at the lower observed equivalent width limit of W $>0.2$\AA.  The absorption line measurements included in this paper have therefore all been drawn from this DEIMOS spectrum.  
\vspace{0.5cm}

 \begin{table*}[t!]
\centering
\begin{minipage}{0.95\textwidth}
\begin{center}
\caption{Ions Detected in Absorption Along the Quasar Sight Line\label{tbl-abslines}}
\begin{tabular}{rrrrrrrrrr}
\tableline\tableline
z$_{abs}$ & $\lambda_{obs}$ & $W_{r}$ & $\sigma_{W_{r}}$ & $\lambda_{r}$ & Ion & z-z$_{abs}$ & $\Delta v$\\
 & [\AA] & [\AA] & [\AA] & [\AA] & &  & [km s$^{-1}$] \\

\tableline
\hline
1.24073 &   5339.77   & 0.49 &  0.06 &  2382.8  & \brittion{Fe}{ II} &   0.00016 &  -17.1 \\
 &   5826.46  & 0.24 &  0.05 &  2600.2 &  \brittion{Fe}{ II} &  -0.00004  & 4.5   \\
 &   6265.85  & 0.79 &  0.08 &  2796.4  & \brittion{Mg}{II} & -0.00012  & 12.7  \\
&   6281.61  & 0.80 &  0.07 &  2803.5 &  \brittion{Mg}{II}  & -0.00023  & 25.1 \\
 \tableline
 1.35689 &   5522.89  & 0.09 &  0.03 &  2344.2  & \brittion{Fe}{ II}  &  -0.00126  & 134.4  \\
 &   5600.13  & 0.18 &  0.04 &  2374.5  &  \brittion{Fe}{ II}   &   0.00126 &  -135.1  \\
 &   5618.17  & 1.33 &  0.07 &  2382.8  & \brittion{Fe}{ II}  &   0.00061  & -65.7   \\
 &   5856.37  & 0.15 &  0.05 &  2484.0  &  \brittion{Fe }{   I}  &   0.00039  & -42.2  \\
 &   6099.43  & 0.65 &  0.09 &  2586.7  & \brittion{Fe}{ II} &   0.00082   & -87.7  \\
&   6130.74  & 1.28 &  0.08 &  2600.2  &  \brittion{Fe}{ II}  &   0.00060  & -64.0  \\
&   6592.84 &  2.64 &  0.07 &  2796.4 &\brittion{Mg}{II} &  0.00043  & -46.3  \\
&   6609.79  & 2.51 &  0.08 &  2803.5 & \brittion{Mg}{II}  & 0.00044 &  -47.4  \\
\tableline
1.74432 &   5562.73 &  0.33 &  0.03 &  2026.5  & \brittion{Mg}{ I}  &   0.00071 &   -71.8   \\ 
 &   6433.55 &  0.41 &  0.05 &  2344.2  & \brittion{Fe}{ II}  &   0.00012  &  -12.0  \\ 
&   6539.17 &  0.73 &  0.07 &  2382.8  & \brittion{Fe}{ II}  &   0.00004 &  -4.3 \\
&   7072.05 &  0.13 &  0.03 &  2576.9 &  \brittion{Mn}{II}  &  0.00011 &  -10.9  \\ 
 &   7098.18 &  0.31 &  0.05 &  2586.7  & \brittion{Fe}{ II}  &   -0.00016 &  16.3  \\ 
 &   7135.74 &  0.72 &  0.07 &  2600.2 & \brittion{Fe}{ II} & 0.00001  & -1.3  \\ 
 &   7673.88 &  1.82  & 0.09 &  2796.4  & \brittion{Mg}{II} &  -0.00008 &  7.6   \\ 
 &   7693.96 &  1.32 &  0.08 &  2803.5 &  \brittion{Mg}{II} &  0.00006 &  -6.3 \\ 
\tableline \tableline
\tableline
\hline
 0.83056$^{a}$ & 5118.97 & 0.24 & 0.06 & 2796.4 & \brittion{Mg}{II} & - & - \\
\tableline
 0.85515$^{a}$ &  5187.75 &  0.28 & 0.10 & 2796.4 & \brittion{Mg}{II} & - & - \\
\tableline
\footnotetext[1]{Cases in which the equivalent width of \brittion{Mg}{II} ($\lambda$2796) falls below the 4$\sigma$ criterion for independent detection of a reliably identified absorption system.  These absorption features have been measured after first detecting a galaxy with a precise grism redshift and an impact parameter $<$150 kpc.} 		
\end{tabular}
\end{center}
\end{minipage}
\end{table*}

\section{Analysis}

\subsection{Identification of Quasar Absorption Lines}

With the aim of producing a complete set of metal absorption lines in the spectrum of quasar SDSS J123622.93+621526.6, we apply an automated approach similar to the pipeline used to extract metal absorption lines from the SDSS DR7 \citep{York05,L09,York12}.   This approach fits a pseudo-continuum to the quasar spectrum using a moving average with an adaptive smoothing length that depends on the proximity to emission lines in the quasar spectrum.  This technique reliably fits the shape of quasar continua for a wide range in spectral morphologies, with the exception of broad absorption line quasars and regions of the Lyman-$\alpha$ forest.

After normalizing the DEIMOS spectrum by the determined continuum, residual narrow absorption features are fit with a Gaussian function to calculate the equivalent width and accompanying error of each line.  All absorption features with an equivalent width detection significance of W/$\sigma_{W}>$ 3 are retained for identification.  As the doublets of \brittion{Mg}{II} ($\lambda\lambda$2796,2803) and \brittion{C}{IV} ($\lambda\lambda$1548,1551) are prolific in optical spectra over a wide range in redshift and also easily identified by their wavelength separation and doublet ratio, these lines are the first target of our automated line identification algorithm.  After producing a list of candidate doublets from the detected lines, we then determine the redshift of each doublet and proceed to search for other common line transitions detected in absorption within 150 km s$^{-1}$ of the same redshift.

In quasar SDSS J123622.93+621526.6, this automated line detection algorithm identifies three absorption systems with detections of \brittion{Mg}{II} doublets and absorption from two or more additional accompanying metal transitions with $\Delta$v $<150$ km s$^{-1}$ .  The redshifts of the absorption line systems ($z=1.241$, 1.353, 1.749) are determined by averaging the individually determined redshifts of all lines with W/$\sigma_{W}\geq$4.  A full listing of the absorption line measurements is provided in Table ~\ref{tbl-abslines}.

\begin{table*}[t!]
\centering
\begin{minipage}{0.95\textwidth}
\begin{center}
\caption{Galaxies Identified in 3D-HST within 150 kpc of the Quasar Sight Line\label{tbl-emlines}}
\begin{tabular}{rrrrrrrrrrrrr}
\tableline\tableline
ID & 3D-HST$^{a}$ & RA & Dec & z$_{gal}$ & sep & $\rho$& (z$_{gal}$-z$_{abs}$) & A$_{v}$ & log(M$_{*}$)$^{b}$ & log(SFR)$^{b}$ \\
& & [deg] & [deg] &  & [\arcsec] &  [kpc] &  & [mag] & [M$_{\odot}$] & [M$_{\odot}$yr$^{-1}$] \\

\tableline\tableline
\hline
G$-$1 & 10245 &  189.09701171 & 62.25773284 & 1.356 & 2.72 & 22.8 & -0.0004 & 0.9 & 10.02$\pm$0.10 & 1.20$\pm$0.12\\
G$-$2 & 10167 & 189.09505985 & 62.25677527 & 1.749 & 2.30 & 19.5 & 0.0043 & 0.5 & 9.44$\pm$0.10 & 1.22$\pm$0.13\\
G$-$3 & 10225 & 189.09169228 & 62.25763867 &1.239 & 6.62 & 55.2 & -0.0013 & 0.5 & 9.54$\pm$0.05 &1.28$\pm$0.05\\
G$-$4 & 10138 & 189.09873233 & 62.25622073 & 0.959 & 6.67 & 52.9 & - & 0.4 &  9.12$\pm$0.10 & 0.33$\pm$0.14 \\
G$-$5 & 10007 & 189.10030524 & 62.25449792 & 0.831 & 12.98 & 98.7 & 0.0002 & 0.4 & 9.25$\pm$0.19 & 0.33$\pm$0.05\\
G$-$6 & 09913 & 189.10052481 & 62.25325999 & 0.857 & 16.92 & 129.9 & 0.0023 & 0.1 &10.06$\pm$0.05 & $-$0.92$\pm$0.05\\
G$-$7 & 10021 & 189.10602084 & 62.25540397 & 0.656 & 18.83 & 131.0 & - & 0.2  &10.05$\pm$0.18 & 0.56$\pm$0.05 \\
G$-$8 & 10152 & 189.10132430 & 62.25699568 & 0.683 & 9.68 & 68.5 & - & 0.9 &10.72$\pm$0.10 & 0.15$\pm$0.05\\

\tableline
\footnotetext[1]{The unique 3D-HST identification number, which formally includes a prefix of ``GOODS-N-14-" referring to the field and pointing.}
\footnotetext[2]{Stellar masses and SFRs estimated assuming the best-fit G141 redshift and SPS modeling of SEDs with 11-15 broad-band flux measurements. The SPS modeling assumes a \citet{Salpeter55} initial mass function and stellar population models from \citet{bc03}. The quoted error estimates are derived exclusively from the photometric uncertainties and do not account for the larger systematic uncertainties from the choice of SPS model or initial mass function.} 			
\end{tabular}
\end{center}
\end{minipage}
\end{table*}

 \begin{table*}[]
\centering
\begin{minipage}{0.85\textwidth}
\begin{center}
\caption{Galaxy Emission Line Measurements in G141 \label{tbl-grism}}
\begin{tabular}{rrrrrrrrr}
\tableline\tableline
ID &  z$_{gal}$ &  W$_{r}$(\brittion{H}{$\beta$})& W$_{r}$(\brittion{[O}{III]}) & W$_{r}$(\brittion{H}{$\alpha$}) & F(\brittion{H}{$\alpha$}) & SFR$(H\alpha)$  &$\Sigma$SFR$(H\alpha)$\\
&  & [\AA]  & [\AA]  & [\AA] & [10$^{-17}$ergs s$^{-1}$ cm$^{-2}$] & [M$_{\odot}$yr$^{-1}$] & [M$_{\odot}$yr$^{-1}$kpc$^{-2}$] \\
\tableline\tableline
\hline
G$-$1 & 1.356 & 6.7$\pm$12.7 & 18.8$\pm$22.3 &  205.5$\pm$28.7 & 8.25$\pm$0.90 & 7.25$\pm$0.79 &  0.37$\pm$0.05\\
G$-$2 & 1.749 & 189.3$\pm$33.1  & 536.6$\pm$57.4 & - & -& - & - \\
G$-$3 & 1.239 & 1.4$\pm$4.1 & 43.4$\pm$13.1 & 159.3$\pm$13.7 & 6.52$\pm$0.50  &  4.57$\pm$0.35 &  0.98$\pm$0.12\\
G$-$4 & 0.959  & - & - &  $<$6.1 & 0.0 & - & -\\
G$-$5 & 0.831 & - & - & 96.6$\pm$50.0 & 4.69$\pm$2.49 & 1.22$\pm$0.65 & 0.02$\pm$0.01 \\
G$-$6 & 0.857 & - & - & $<$5.9 & 0.0 & - & -\\ 
G$-$7 & 0.656 & - & - & 79.4$\pm$34.5 & 16.54$\pm$7.97 & 2.42$\pm$1.16 & 0.05$\pm$0.02\\
G$-$8 & 0.683 & - & - & 72.6$\pm$9.4 & 22.79$\pm$2.94 & 3.68$\pm$0.47 & 0.16$\pm$0.02 \\
\tableline
\end{tabular}
\end{center}
\end{minipage}
\end{table*}

\begin{figure*}[t!]
\centering
\begin{center}
\epsscale{0.92}
\plotone{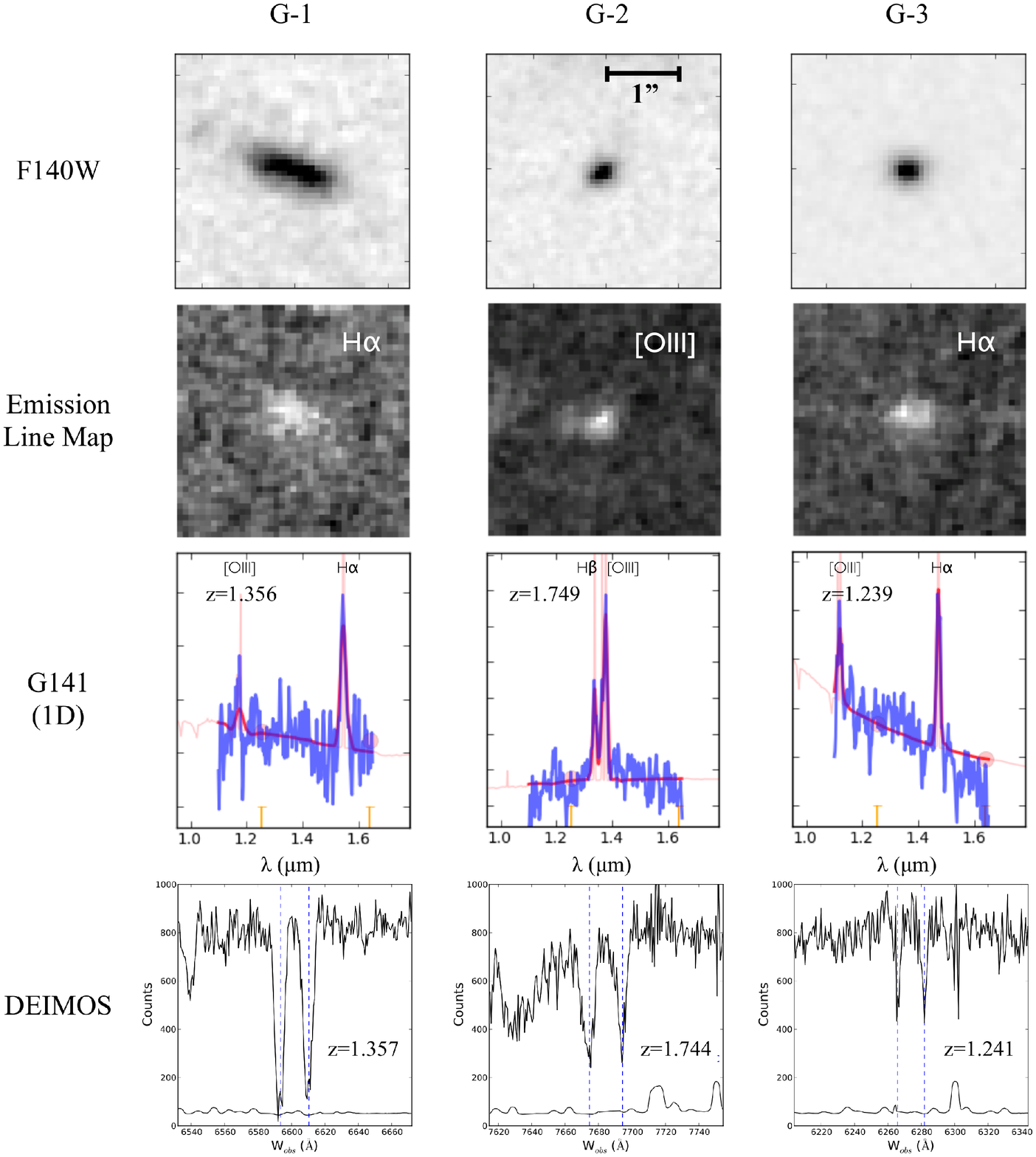}
\caption{{\it Top row:} WFC3 F140W 2\arcsec x2\arcsec~imaging cutouts for each of the three star-forming galaxies labeled in Figure~\ref{fig:fov} as G$-$1, 2, \& 3.  For physical scale, 1\arcsec~is equivalent to 8.45 kpc at the mean redshift of this sample, $z=1.44$. {\it Second row:} Emission line maps for each galaxy, produced using a combination of the WFC3 F140W imaging and G141 2D grism data.  {\it Third row:}  WFC3 G141 1D grism observations are shown in blue with a best-fit template galaxy spectrum overlaid in red.  Best-fit redshifts are inset.  {\it Bottom:}  The corresponding three portions of the KTRS DEIMOS spectrum and 1$\sigma$ error array of the nearby quasar  SDSS J123622.93+621526.6, in which a \brittion{Mg}{II} absorption doublet is detected with a redshift consistent with that of each galaxy in WFC3 G141, at projected separations of $<60$ kpc.  The grism redshift of each galaxy precisely matches the mean redshift of a detected \brittion{Mg}{II} absorption line system within $\Delta z\leq0.005$.  \label{fig:galfig}}
\end{center}
\end{figure*}

\subsection{Analysis of Proximate Galaxies}

Having identified all significant absorption features along the line of sight to our central quasar, we next examine the WFC3 F140W and G141 observations to identify all galaxies with $0.65<z<2.6$ in the surrounding field.   The applied redshift limitations correspond to the range in which emission from either \brittion{H}{$\alpha$} or the \brittion{[O}{III]}/\brittion{H}{$\beta$} complex is observable in the G141.  Although the 5$\sigma$ depth of the F140W images extends to $H_{140}\sim26$, the continuum limit at which grism redshifts may be reliably obtained for galaxies at $1<z<3$ in the 3D-HST Survey effectively limits our search to galaxies with $H_{140}\lesssim24$ \citep{Brammer12}. We stress that in this search we pay no regard to the absorption properties of the galaxies.  {\it Thus, our analysis can be used interchangeably as a galaxy-blind survey of absorbers with W$_{obs}>0.4$\AA~ as well as an absorption-blind survey of galaxies with log(M$_{*}$/M$_{\odot}$) $>9$.}

To our detection limits, eight galaxies are identified within 150 kpc of the quasar sight line, and four galaxies are detected with projected physical separations of less than 60 kpc (see Table ~\ref{tbl-emlines}).  Three of these four closest galaxies exhibit strong emission lines in the extracted 1D spectra from the G141 grism observations, enabling their redshifts to be precisely determined at $z=1.239$, $z=1.356$, and $z=1.749$.  In each of these cases, multiple emission lines are detected in the G141 spectra, ensuring that these redshifts are unambiguous.  Equivalent width measurements of the \brittion{H}{$\alpha$}, \brittion{H}{$\beta$}, and [\brittion{O}{III}] emission lines, where observable are presented in Table ~\ref{tbl-grism}. 

The extracted G141 1D spectra of these three galaxies are presented in Figure~\ref{fig:galfig}, with labels G1$-$3, matching the labels in the F140W image from Figure~\ref{fig:fov}.  In each case, the galaxy emission redshift is found to precisely match the redshift of one of the separately identified absorption systems detected in the quasar spectrum.  Beneath each grism observation in Figure~\ref{fig:galfig}, we present a cutout from the TKRS DEIMOS quasar spectrum, centered on the detected doublet of \brittion{Mg}{II} absorption matching the redshift of the galaxy separately observed in emission.  The difference between the emission and absorption redshifts is $\Delta z\leq0.005$ in each case, indicative of the precision of grism redshifts among star-forming galaxies in the 3D-HST survey.

\subsubsection{Morphologies}

The high resolution and small native pixel scale (0.06\arcsec/pix) of the WFC3 F140W images further enable us to examine the morphologies of the galaxies proximate to the quasar sight line.  Using the {\it GALFIT} software package \citep{Peng02}, we derive the effective radius (R$_{e}$), S{\'e}rsic index, axis ratio (b/a), and position angle (PA) for each of the eight galaxies identified within 150 kpc of the quasar sight line.  These measurements and errors, both from {\it GALFIT}, are presented in Table ~\ref{tbl-galfit}.

From the measured position angles, we also calculate the azimuthal angle, $\Phi$, for each galaxy, relative to the quasar.  We define this angle using the same convention as \citet{Kacprzak12} and \citet{B12b}, in which $\Phi$ refers to the angle produced by the major axis of the galaxy and the vector relating the center of the galaxy to that of the quasar.  For cases in which the quasar is aligned with the major axis of the galaxy, $\Phi=0^{\circ}$, and $\Phi=90^{\circ}$ when the quasar aligns with the minor axis of the galaxy.

The slitless nature of the WFC3 grism also enables the extraction of 2D emission line maps, as recently demonstrated by \citet{Nelson12}.  We use the same technique of fitting and subtracting the 1D continuum emission, row by row, from the 2D spectrum to produce emission line maps for the three galaxies closest to the quasar sight line.  Such maps for the three galaxies observed in absorption at $z>1$ are provided in Figure~\ref{fig:galfig}.  For the two galaxies with observable \brittion{H}{$\alpha$} in the G141 spectrum we find that the resolved emission line regions are extended, allowing us to rule out a dominant flux contribution from an AGN in each case.  We also note that the \brittion{H}{$\alpha$} emission generally traces the rest-frame optical morphology of each galaxy, consistent with the findings of \citet{Nelson12}.

\subsubsection{Stellar Population Estimates}

For each of the galaxies with \brittion{H}{$\alpha$} emission line measurements in the G141 spectra we estimate the SFR using the relation from \citet{Kennicutt98}:
\begin{eqnarray}
\textrm{SFR}(\brittion{H}{$\alpha$})[M_{\odot}/\textrm{yr}] = 7.9 \times 10^{-42} \times L(\brittion{H}{$\alpha$})[\textrm{erg/s}]
\label{eqn-1}
\end{eqnarray}
and present the measurements in Table ~\ref{tbl-grism}.  We apply no correction for dust extinction, so these estimated SFRs should be understood as lower limits of the true SFR.


For galaxies in which strong emission lines are not evident or observable in the G141 1D spectra, ancillary multi-band photometric observations facilitate the estimation of the stellar masses and SFRs from stellar population synthesis (SPS) modeling.  The stellar masses and SFRs, estimated from the SED of each galaxy alone, are given in Table ~\ref{tbl-emlines}.  These estimates are obtained using the FAST code \citep{Kriek09}, assuming a \citet{Salpeter55} initial mass function and the stellar population models of \citet{bc03}.  We also assume an exponentially declining ($\tau$ model) star formation history and solar metallicity.  In estimating the SFRs, we apply the prescriptions of \citet{Wuyts11}, by requiring a minimum e-folding time of log($\tau_{min})=8.5$, which has been shown to best reproduce the low-to-intermediate SFRs within the \citet{bc03} framework.  For the estimation of stellar masses we allow more freedom in the best-fit SPS models, setting log($\tau_{min})=7$ \citep[see also][submitted]{L12a}.   

We note that the SFR estimates derived from SPS modeling from {\it FAST} are consistently $>2$ times larger than the SFRs extracted from the H$\alpha$ emission line measurements.  This difference, which is consistent with the findings of  \citet{Nelson12}, can be attributed to the fact that the SPS modeling simultaneously fits for dust extinction, whereas a similar correction has not been made to the SFRs derived from the grism data.

\begin{table*}[t!]
\centering
\begin{minipage}{0.95\textwidth}
\begin{center}
\caption{Galaxy Morphological Parameters in F140W \label{tbl-galfit}}
\begin{tabular}{rrrrrrrrr }
\tableline\tableline
ID &  m$_{F140}$ &  R$_{e}^{a}$ & R$_{e,c}^{b}$ & $n^{c}$ & b/a & PA & $\Phi^{d}$ \\
& & [pix] & [kpc] & &  & [deg] & [deg]\\
\tableline\tableline
\hline
G$-$1  & 23.15 & 6.35$\pm$0.25 & 1.77$\pm$0.11 & 0.50$\pm$0.06 & 0.26$\pm$0.01 & 69.2$\pm$1.0 & 37$\pm$2  \\
G$-$2  & 23.93 & 2.23$\pm$0.25 &  0.70$\pm$0.14 & 1.75$\pm$0.45 & 0.33$\pm$0.05 & -50.3$\pm$3.2 & 46$\pm$14\\
G$-$3 & 23.07 & 1.93$\pm$0.11 &  0.86$\pm$0.08 & 2.68$\pm$0.43 & 0.68$\pm$0.05 & 87.5$\pm$5.6 & 17$\pm$6  \\
G$-$4 & 23.59 & 2.82$\pm$0.12 &  1.15$\pm$0.08 & 1.05$\pm$0.18 & 0.63$\pm$0.04 & -68.6$\pm$4.9 & - \\
G$-$5 & 23.35 & 7.37$\pm$0.71 &  2.81$\pm$0.36 & 0.94$\pm$0.14 & 0.60$\pm$0.04 & -21.9$\pm$5.5 & - \\
G$-$6 & 22.45 & 1.69$\pm$0.13 &  0.72$\pm$0.07 & 6.84$\pm$1.13 & 0.74$\pm$0.04 & -15.9$\pm$5.3 & - \\
G$-$7 & 21.64 & 9.03$\pm$0.26 &  2.68$\pm$0.11 & 1.73$\pm$0.06 & 0.43$\pm$0.01 & -3.9$\pm$0.7 & - \\
G$-$8 & 21.33 & 4.89$\pm$0.04 &  1.89$\pm$0.03 & 0.82$\pm$0.02 & 0.71$\pm$0.01 & 31.6$\pm$1.0 &- \\
\tableline
\footnotetext[1]{R$_{e}$ indicates the effective radius of the galaxy light profile along the semi-major axis.  }
\footnotetext[2]{The circularized effective radius, R$_{e,c}$=R$_{e}\sqrt{b/a}$, where (b/a) is the axis ratio, provided in the fifth column of the table.}
\footnotetext[3]{The S{\'e}rsic index, estimated using the {\it GALFIT} software \citep{Peng02}.}
\footnotetext[4]{Azimuthal angle to the quasar sight line.} 	

\end{tabular}
\end{center}
\end{minipage}
\end{table*}

Using the \brittion{H}{$\alpha$}-derived SFRs when available, we calculate the average star formation surface density, $\Sigma$SFR, as:
\begin{eqnarray}
\Sigma\textrm{SFR} [M_{\odot}\textrm{yr}^{-1} \textrm{kpc}^{-2}] = 0.5\times \textrm{SFR} / \pi R_{e,c}^{2}
\label{eqn-2}
\end{eqnarray}
where R$_{e,c}$ is the circularized effective radius of the galaxy in the F140W image, defined as:
\begin{eqnarray}
R_{e,c} = R_{e} \sqrt{b/a}
\label{eqn-3}
\end{eqnarray}
Here, R$_{e}$ is the effective radius along the semi-major axis measured in kpc, and $(b/a)$ is the axis ratio, described in Section 3.2.1.  These measurements are also presented in Table ~\ref{tbl-grism}.  While circularizing the effective radius would be inappropriate for thin disks, in which inclination has a large effect on the measured surface area, we find it to be a safe approximation for the more elongated, irregular, and puffy morphologies of the galaxies in our analysis.


\section{Discussion}

Our essentially unbiased search for the galaxy hosts of \brittion{Mg}{II} absorption with W$_{r}^{\lambda2796}>0.4$\AA~in the 3D-HST Survey area produces a 100\% detection rate in the available redshift range of the G141 grism.  In each case, the \brittion{Mg}{II} host galaxies are located within 60 kpc of the quasar sight line and have resolved emission lines and SEDs implying SFRs $>4$ M$_{\odot}$ yr$^{-1}$.  All of the galaxies appear to be isolated with no evidence of recent disruption.  Thus there is also no ambiguity in the absorber-galaxy pairing due to multiple candidate galaxy hosts at similar redshifts.

\subsection{Properties of \brittion{Mg}{II} Host Galaxies}

For the three galaxies at $z>1$ exhibiting strong \brittion{Mg}{II} absorption in the quasar spectrum, we measure effective radii along the semi-major axis for these galaxies ranging from 1$-$3 kpc and S{\'e}rsic indices ranging from $0.5\lesssim n\lesssim 2.7$.   We note that the true uncertainties in the structural parameter estimates are larger than those reported by {\it GALFIT} \citep[e.g.,][]{Haussler07}.  If we assume that our small but randomly-selected sample is representative of the larger \brittion{Mg}{II} host galaxy population at $1<z<2$, our data suggest that \brittion{Mg}{II} traces compact, triaxial systems (e.g., G$-$2, G$-$3) and thick disks (e.g., G$-$1), each of which appear to be common among low-mass (log(M$_{*}$/M$_{\odot}$) $\lesssim10$) galaxies at $z\sim2$ \citep[e.g.,][]{Conselice05, Elmegreen05, Elmegreen06, Ravindranath06, Genzel06, Genzel08, Law12a}. 


The two \brittion{Mg}{II} host galaxies for which \brittion{H}{$\alpha$} is observable in the G141 spectrum (G$-$1 and G$-$3) exhibit SFRs in the range 4$-$8 M$_{\odot}$ yr$^{-1}$ and star formation surface densities in the range 0.3$-$1.0 M$_{\odot}$ yr$^{-1}$ kpc$^{-2}$.  Even without applying a dust correction, which would effectively raise the SFRs, we measure a $\Sigma$SFR for each galaxy that is well-above the threshold over which large-scale outflows are observed in local starburst galaxies and Lyman break galaxies at $z>2$ \citep[0.1 M$_{\odot}$ yr$^{-1}$ kpc$^{-2}$;][]{Heckman02}.  Given that outflow velocities exhibit a stronger correlation with $\Sigma$SFR than with SFR \citep{Kornei12}, it is perhaps not surprising that we find evidence of winds extending to large scales around each of these galaxies.  If we assume a wind velocity of 400 km s$^{-1}$, typical of the observations of star-formation-driven winds at $z\sim1.4$ \citep{Weiner09}, the impact parameters of the galaxies with respect to the quasar sight line imply that the winds were launched at least $\sim50$ and $\sim150$ Myr earlier.  The fact that each of the galaxies is still forming stars at a high rate suggests that we may be observing a prolonged burst of star formation on these same timescales. 

SED modeling of the galaxy G$-$2, which is observed in absorption but at too high a redshift for H$\alpha$ detection in the G141 spectrum, indicates a SFR approximately equal to that of G$-$1 and G$-$3.  We therefore estimate the $\Sigma$SFR to be $\sim2$ M$_{\odot}$ yr$^{-1}$ kpc$^{-2}$, again well above the same starburst threshold.  The \brittion{[O}{III]}/\brittion{H}{$\beta$} ratio for this galaxy from the 1D G141 spectrum is measured to be 2.83$\pm^{0.98}_{0.68}$, consistent with a normal star-forming galaxy.  As shown in Figure~\ref{fig:galfig}, the resolved \brittion{[O}{III]} emission is extended, tracing the rest-frame optical morphology and thus disfavoring a dominant flux contribution from an AGN.  

The three \brittion{Mg}{II} host galaxies we detect at $1<z<2$ have a mean stellar mass of log(M$_{*}$/M$_{\odot}$) $=9.75$.  Using the relation between stellar mass and halo mass in this redshift range from \citet{Wake11}, this corresponds to a dark matter halo mass of log(M$_{h}$/M$_{\odot}$) $\sim11.9$.  We therefore find consistent results with the typical W$_{r}>0.8$\AA~ \brittion{Mg}{II} halo mass derived from clustering measurements at $z\sim1$, 1.8$\pm^{4.2}_{1.6}\times$10$^{12}$M$_{\odot}$ \citep{L11}.  This value is also consistent with multiple clustering measurements at $z\sim0.6$ \citep{B06, L09, Gauthier09}, indicating that \brittion{Mg}{II} host galaxies occupy similarly massive haloes at all redshifts $z\lesssim2$. 

\begin{figure*}[t!]
\centering
\begin{center}
\epsscale{0.85}
\plotone{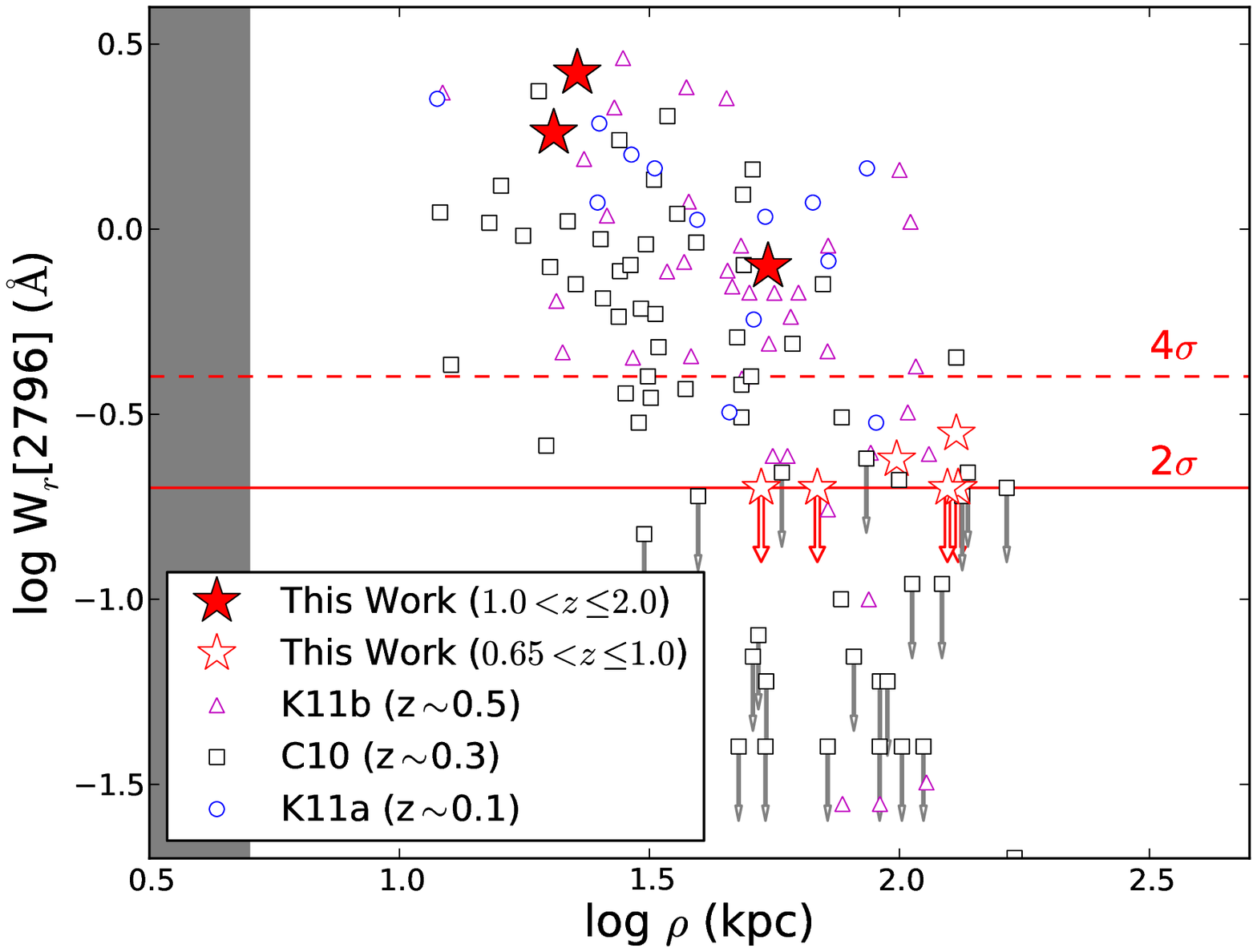}
\caption{The observed relation between the rest-frame equivalent width of \brittion{Mg}{II} absorption, W$_{r}^{\lambda2796}$, and galaxy impact parameter, $\rho$ in this work and from the largest existing surveys.  Measurements from the literature are overplotted for comparison at low and intermediate redshift (circles, \citet{Kacprzak11a}; squares, \citet{Chen10a}; triangles, \citet{Kacprzak11b}).  The shaded gray region indicates the limiting observable impact parameter for this work.  Red horizontal lines provide the significance of absorption line detection in the DEIMOS quasar spectrum we examine.  \label{fig:newchen}}
\end{center}
\end{figure*}

Despite our unavoidably small sample, a 100\% detection rate of star-forming galaxy hosts of W$_{r}^{\lambda2796}>0.8$\AA~\brittion{Mg}{II} absorption at $z>1$ may be interesting given the results from recent ground-based analyses, which have reported a large fractional absence of strong \brittion{Mg}{II} host galaxy detections to sensitive limits in the same redshift range.  A search for the galactic counterparts of 20 ultra-strong (W$_{r}^{\lambda2796}>2$\AA) \brittion{Mg}{II} absorbers at $z=2$ using SINFONI undertaken by \citet{B12a} had only 20\% success in identifying absorber host galaxies, at a reported SFR sensitivity limit of 2.9 M$_{\odot}$yr$^{-1}$.   This detection rate represents a drop from 66\% at $z\sim1$ with a similar search \citep{B07}.  The SINFONI field of view limits a search for companions at $z=2$ to impact parameters less than $\sim40$ kpc, but the substantial observational evidence indicating that ultra-strong \brittion{Mg}{II} absorbers are restricted to small impact parameters implies that the missing galaxies are not simply beyond the field of view.  \citet{B12a} also rule out scenarios in which the galaxy hosts are obscured by the quasar at very small impact parameters.  Thus, the authors hypothesize that the SFRs of W$_{r}>2$\AA~\brittion{Mg}{II} host galaxies may be substantially lower than expected at $z\sim2$, causing them to lie below the detection threshold of the survey.

The strongest \brittion{Mg}{II} absorber in our sample (G$-$1, $z=1.357$) meets the definition for ultra-strong absorbers studied in the analyses of \citet{B07} and \citet{B12a}.  With a projected separation of 22.8 kpc and a SFR $\sim7$ M$_{\odot}$yr$^{-1}$, the host galaxy we detect is also within the field-of-view and SFR limitations of SINFONI observations and would most certainly have been detected by the H$\alpha$ surveys of Bouch{\'e} et al.  The second-strongest absorber in our sample (G$-$2, $z=1.749$) has a \brittion{Mg}{II} equivalent width of W$_{r}=1.82\pm0.09$\AA, slightly weaker than the classic definition of ultra-strong \brittion{Mg}{II}.  Still, the projected separation between the host galaxy and the quasar sight line is 19.5 kpc, which is again well within the SINFONI field of view, and the estimated SFR is also above the detection threshold of the SINFONI observations.  While our success rate at detecting these host galaxies at small impact parameters hints at a contradiction with the low detection rates from Bouch{\'e} et al., a sample of two objects cannot provide optimal constraints on the typical SFRs of \brittion{Mg}{II} host galaxies at $z>1$.  A greater number of G141 observations centered on quasar sight lines with high redshift \brittion{Mg}{II} absorption would be required to better examine whether lower-than-predicted SFRs can account for the large number of host galaxies undetected in other surveys. 

\begin{figure*}[ht!]
\centering
\epsscale{0.8}
\plotone{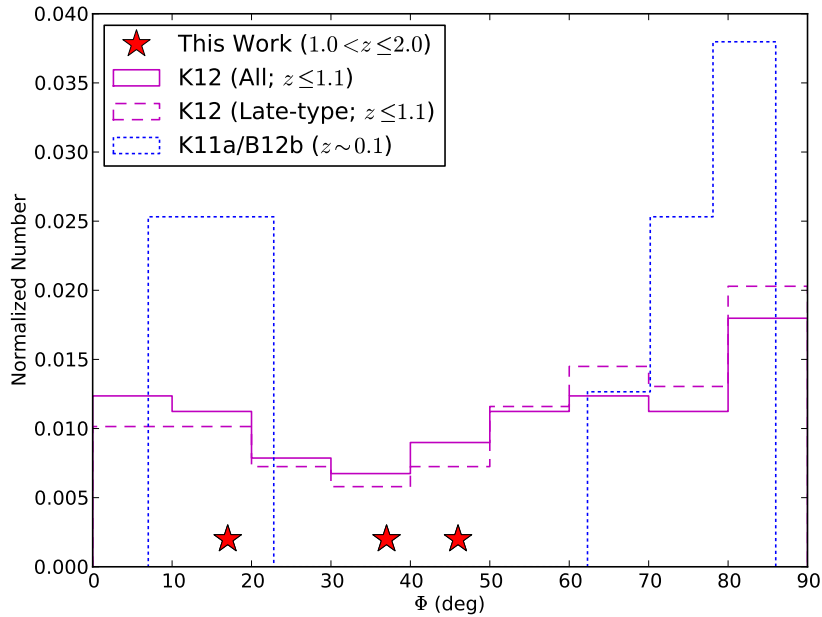}
\caption{The distribution of the azimuthal angle ($\Phi$) of galaxies, relative to the quasar sight lines in which \brittion{Mg}{II} absorption is detected at W$_{r}>0.1$\AA.  We compare our three individual high redshift measurements of $\Phi$ (stars) with an ensemble of observations from two surveys at lower redshift.  The lowest redshift data from \citet{Kacprzak11a} and \citet{B12b} consist of 11 galaxy-absorber pairs at $z\sim0.1$.  The intermediate redshift data from \citet{Kacprzak12} includes 88 galaxy-absorber pairs drawn from a compilation of studies \citep{Chen10a, Kacprzak11a, Kacprzak11b, Churchill12},  shown in full (solid) and for late-type galaxies only (dashed).  The data hint at an evolving azimuthal distribution of gas contained in outflows and traced by \brittion{Mg}{II}, which is consistent with the observed evolution in the collimation of outflows from star-forming galaxies from $z\sim3$ \citep{Law12}.  \label{fig:azdist}}
\centering
\end{figure*}


While this work is primarily focused on galaxies with z$>$1, we identify two galaxies at lower redshift with impact parameters of $\sim100$ kpc (G$-$5, $z=0.831$; G$-$6, $z=0.856$) and corresponding 3$\sigma$ detections of \brittion{Mg}{II} with W$_{r}\sim0.25$\AA.  We note that due to the small equivalent widths of the \brittion{Mg}{II} 2796\AA~ transition in each case, no additional absorption lines are detected above the 3$\sigma$ threshold at the same redshift, making the absorption identifications less certain than the higher equivalent width \brittion{Mg}{II} systems previously discussed.  However, both of these systems exhibit an absorption line with 1$\sigma$ significance at the expected location of the 2803\AA~ transition of the \brittion{Mg}{II} doublet, indicating that despite the weakness of the features the detections are real.  As shown in Figure ~\ref{fig:newchen}, the equivalent widths and impact parameters measured for these absorbing galaxies also agrees well with the previously determined relations at lower redshift.

The galaxy identified as G$-$5 has a well-identified H$\alpha$ emission line in its G141 spectrum, with a luminosity implying a SFR $=1.22$ M$_{\odot}$yr$^{-1}$.  Galaxy G$-$6 exhibits no observable emission lines, despite having a well-determined photometric redshift.  The SPS modeling of its SED implies a SFR $\sim0.1$ M$_{\odot}$yr$^{-1}$, consistent with the lack of H$\alpha$ emission.  Both galaxies fall well below the $\Sigma$SFR threshold believed to launch large-scale winds.  Given the large projected distances at which the \brittion{Mg}{II} is observed around these galaxies, it is possible that the enriched gas was launched during a burst of star formation as much as 300 Myr earlier.  

While the lower redshift absorption features are not the focal point of this work, it is interesting to note that their \brittion{Mg}{II} host galaxies, which coincidentally are detected at larger impact parameters ($\sim$100 kpc) relative to the quasar sight line, exhibit a seemingly wider range in SFRs but similar stellar masses.  Numerous studies at $z\lesssim1$ have reported the detection of \brittion{Mg}{II} around galaxies with a diverse range of morphological and spectral types \citep[e.g.,][]{Steidel94}, including quiescent galaxies, similar to G$-$6 \citep[e.g.,][]{GC11}.  Hence, the apparent host diversity we find is not unusual.

\subsection{The Spatial Distribution of \brittion{Mg}{II} Around Galaxies}

The inverse relationship between \brittion{Mg}{II} absorption equivalent width and galaxy impact parameter assumed in the survey design of \citet{B12a} is well-established in the low and intermediate redshift Universe \citep[e.g.,][]{LB90, Chen10a, Chen10b} and has been argued to drive the strong observed correlation between \brittion{Mg}{II} equivalent width and dust extinction in background quasars \citep{Menard08}.  A tight inverse relation of W$_{r}$ and $\rho$ could also explain the strong correlation between \brittion{Mg}{II} equivalent width and [\brittion{O}{II}] emission observed in stacks of SDSS absorption spectra \citep{Menard11}.  

Significant evolution in the observed scaling of the W$_{r}-\rho$ relation could result in the non-detection of \brittion{Mg}{II} host galaxies in surveys with a limited field of view \citep[e.g.,][]{B07,B12a}.  While our data set is small, its unbiased selection and completeness enables us to look for any obvious departures from these scaling relations at lower redshift.  In Figure~\ref{fig:newchen}, we compare our measurements at $z\sim1.5$ to those in the largest existing low and intermediate redshift surveys \citep{Chen10a, Chen10b, Kacprzak11a, Kacprzak11b}.  While our observations appear to skirt the upper edge of the measurements from the absorber-blind survey of \citet{Chen10a,Chen10b}, they agree quite well with the absorption-selected galaxy observations of \citet{Kacprzak11a, Kacprzak11b}.  Thus, we find no signs of dramatic evolution in this relation from $z\sim1.5$.

The large scatter in the W$_{r}-\rho$ relation, which is equivalent to $\sim1$ dex along each axis, is also well-established \citep[e.g.,][]{Chen10a, Chen10b,Kacprzak11b}.  Variance in one or more of the many properties of absorber host galaxies likely lies at the root of this scatter, and recent work by \citet{Chen10b} indicates that the stellar mass of the host galaxy is the fundamental driver.  This fact is surprising, given the substantial body of evidence linking the strongest \brittion{Mg}{II} absorbers to starburst galaxies at $z>1$ \citep[e.g.,][]{B07,Nestor11}, and photometric stacks of residual light around quasars in the SDSS indicate that stronger (W$_{r}>1$\AA) \brittion{Mg}{II} absorbers preferentially trace young, star-forming galaxy populations \citep{Zibetti07}, compared to (W$_{r}<1$\AA) \brittion{Mg}{II} absorbers.

The various recent studies linking strong \brittion{Mg}{II} absorbers to star-forming galaxies at intermediate redshifts suggests that a substantial fraction of these absorbers originate in star formation-driven winds.   As such outflows are expected to propagate parallel to the minor axis of galaxies \citep[e.g.,][]{Strickland04}, a signature of a wind-driven origin is expected in the azimuthal distribution of absorbers around galaxies.  The first compelling evidence of the preferential distribution of strong \brittion{Mg}{II} along the minor axis of star-forming galaxies was reported in the stacked absorption line profiles of intermediate redshift galaxies in zCOSMOS \citep{Bordoloi11}.  Using deep {\it HST} images to resolve the morphologies and position angles of individual \brittion{Mg}{II} galaxies, \citet{Kacprzak11b, Kacprzak12} confirmed a bimodal azimuthal distribution of \brittion{Mg}{II} absorption in the halos of $z\sim0.5$ host galaxies, finding that the bulk of these absorbers are aligned within 20$^{\circ}$ of the major or minor axis of the host.  Investigating a smaller sample of 11 galaxies, \citet{B12b} report an even more striking bimodality at z$\sim$0.1.

To investigate the persistence of this azimuthal dependence at higher redshift, we examine the spatial distribution of the \brittion{Mg}{II} absorption around the $z>1$ galaxies in this work.  In Figure~\ref{fig:azdist}, we compare our measurements to the azimuthal distributions of absorber-galaxy pairs with W$_{r}>0.3$\AA~ from \citet{B12b} and W$_{r}>0.1$\AA~ from \citet{Kacprzak12}.   While the size of our sample limits what significant conclusions may be drawn from the distribution of $\Phi$ in this analysis, our data are the first measurements of this kind above z$\sim$1 and therefore merit discussion.

At  $1<z<2$, we find that the azimuthal angles ($\Phi$, previously defined in Section 3.2.2) range from 17$^{\circ}-$46$^{\circ}$ with a typical uncertainty of 7$^{\circ}$.   One third of the absorbers in our sample have an azimuthal angle within 20$^{\circ}$ of the major axis, precisely in agreement with the observations of \citet{Kacprzak12}, who suggest that this minority of absorbers can be explained by gas accretion occurring parallel to the major axes of the galaxies.  However, 2/3 of our detections are found with $30^{\circ}<\Phi<60^{\circ}$.  This places our remaining absorbers outside the angular window ($\Phi>60^{\circ}$) within which the bulk of \brittion{Mg}{II} has been shown to inhabit at $z<1$, particularly around star-forming galaxies \citep{Bordoloi11, Kacprzak12} and at low redshift \citep{B12b}.

The collimation of outflows along the minor axis of galaxies observed in local starbursts \citep[e.g.,][]{Heckman90, LH96, Cecil01, SH09} implies a correlation between the inclination of galaxies and their wind velocities, which can be measured using the blueshifts of low-ionization absorption features of galaxies hosting outflows.  This correlation has been observed locally, where galaxy morphologies are readily resolved \citep{Heckman00,YChen10}, and with deep multi-band imaging at $z=1$ \citep{Kornei12}.  However, observations at higher redshift ($z=1.4$) by \citet{Weiner09} failed to reproduce the relation, possibly due to the difficulty of resolving galaxy morphologies at higher redshift, even with {\it HST} {\it I}-band imaging.   A recent study by \citet{Law12} incorporated deep {\it HST} WFC3/IR imaging to better resolve the rest-frame optical morphologies of star forming galaxies at $z>2$.  Their findings indicate that the correlation between outflow velocity and inclination is indeed absent in low-mass galaxies at $z\sim2-3$, suggesting that the typically more irregular and ``puffy'' high redshift galaxies have poorly collimated outflows.  This increasing collimation of star formation-driven winds with redshift is an expected result of galaxies evolving with time from low-mass dispersion-dominated systems to more massive, stable disks.  

If the typical geometry of star formation-driven outflows evolves with time to become highly collimated in the local universe, a signature of this evolution should be expected in the azimuthal distribution of metal-enriched gas in the circumgalactic medium surrounding star forming galaxies.  Despite the small sample size, our data are consistent with an evolutionary trend in which the collimation of galaxy-scale outflows increases with time.  Moreover, the fact that the typical halo masses of \brittion{Mg}{II} host galaxies appears not to evolve from z$\sim$2 suggests that the trend of increasing collimation with time holds for star-forming galaxies at constant mass (i.e., the outflows of star-forming galaxies with log(M$_{*}$/M$_{\odot}$) $=9.75$ are randomly oriented at $z>2$ and more tightly collimated at lower z). We stress that a more extensive survey of \brittion{Mg}{II} host galaxies with {\it HST} would be required to determine the significance of this finding.

\begin{figure*}[ht!]
\centering
\epsscale{0.9}
\plotone{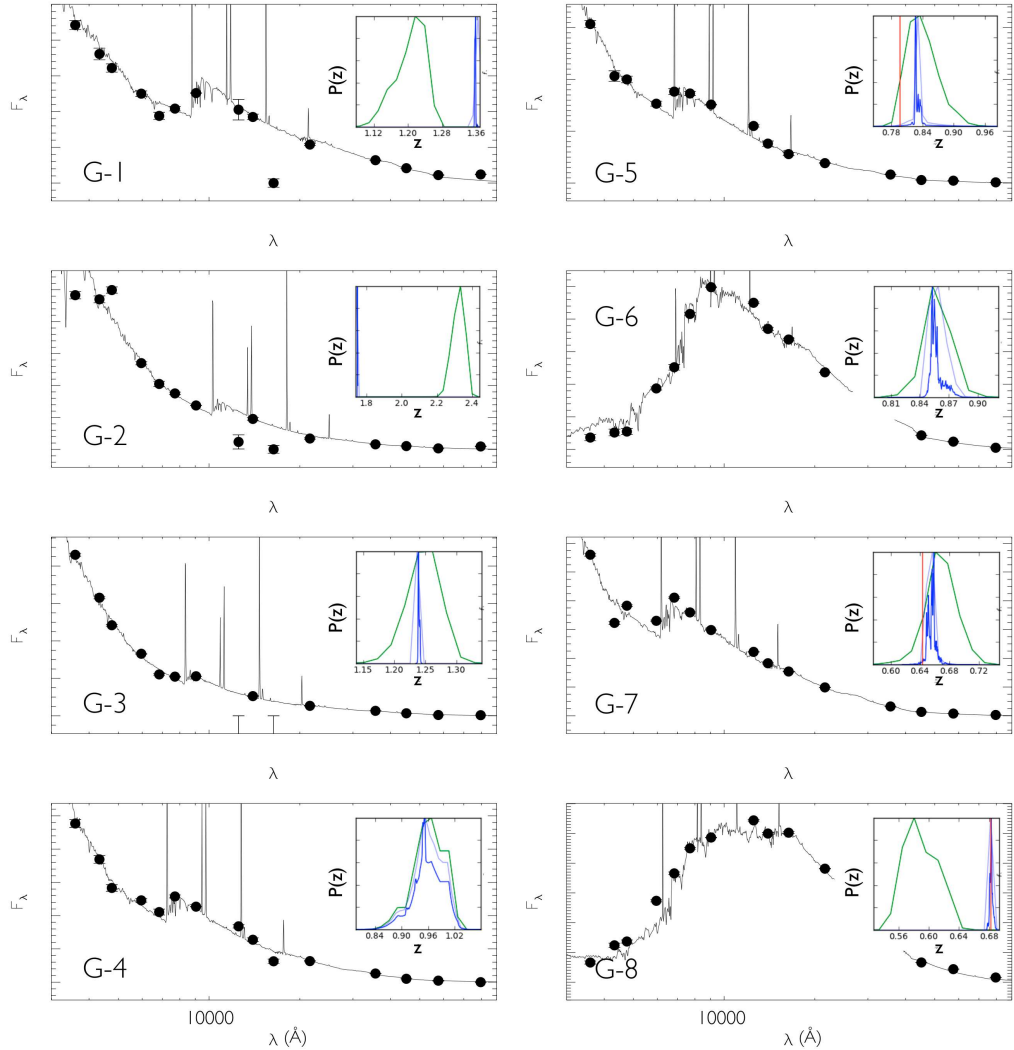}
\caption{The broad-band SEDs of the eight galaxies identified within 150 kpc of the quasar sight line, provided with arbitrary units of $F_{\lambda}$.  In each case the best-fit galaxy template, fit to the redshift determined from the grism spectrum, is overplotted.  Redshift probability distribution functions, $P(z)$, are inset within each panel.  The green curve provides the broad-band photometric solution of the $P(z)$.  The blue curve provides the $P(z)$ determined from the G141 data. \label{fig:galseds}}
\centering
\end{figure*}

\subsection{Absorption Properties of All Proximate Galaxies}

In a parallel search for absorption around all 3D-HST galaxies identified within 150 kpc of the quasar sight line, we detect 8 galaxies with well-determined redshifts in the range $0.65<z<2.6$.  We provide the SED and redshift probability density function, $P(z)$ for each galaxy in Figure~\ref{fig:galseds}. All four of the galaxies with $z>0.9$ are coincidentally located within 60 kpc of the quasar sight line, while the galaxies with lower redshifts are detected at larger impact parameters.   The separations in the redshifts of these objects ensures that there is no physical correlation between any of these galaxies or with the quasar.   Thus, the apparent angular clustering of high redshift galaxies around the quasar sight line should be understood simply as a coincidence and not distract from the other underlying physical correlations of absorbers and host galaxies in this study.

Three of the four galaxies with $\rho<60$ kpc exhibit \brittion{Mg}{II} absorption with W$_{r}>0.8$\AA, as discussed in length in Section 4.1.  The remaining galaxy within this radius is undetected in \brittion{Mg}{II} to our observational limit of W$_{r}>0.4$\AA.   Of the galaxies detected in absorption, all have SFR $>4$ M$_{\odot}$ yr$^{-1}$ and $\Sigma$SFR $>0.3$ M$_{\odot}$yr$^{-1}$kpc$^{-2}$, uncorrected for dust extinction.  The remaining galaxy, G$-$4, which is undetected in absorption at a similarly close separation of 57 kpc, exhibits no measurable emission lines in the G141 data at the best-fit combined grism/photometric redshift of 0.959.  Although we only measure an upper-limit for the \brittion{H}{$\alpha$} emission in the 1D grism spectrum (W$_{r}\sim6$\AA), SPS modeling of the SED results in a SFR $\sim2$ M$_{\odot}$yr$^{-1}$.  

The possibility remains that the best-fit photometric redshift for G$-$4 could be incorrect, which might simultaneously explain the lack of emission lines in the G141 spectra and the absence of \brittion{Mg}{II} absorption in the quasar spectra where they are expected.  However, the redshift probability distribution function derived from the SED of the galaxy indicates a $<$1\% probability that the observable wavelength of H$\alpha$ emission lies outside of the range of the G141 spectrum.  Thus, the fact that we find no strong evidence for rigorous ongoing star formation in the only galaxy within 60 kpc that is undetected in \brittion{Mg}{II} absorption suggests that the SFR of the host galaxy is indeed correlated with the large-scale distribution of cold gas in the circumgalactic medium.  

Taken together, our data indicate that the covering fraction ($f_{c}$) of cold gas with W$_{r}^{2796}>0.8$\AA~ may be as large as unity to at least 60 kpc around star-forming galaxies at $1\lesssim z\lesssim2$.  Given the small number of objects available for this study, we are unable to place tight constraints on these measurements.  However, it is interesting to note that this covering fraction estimate is quite high relative to measurements at lower redshift.  A recent examination of the \brittion{Mg}{II} absorption properties of galaxies in the DEEP2 survey estimated $f_{c}=0.5$ around typical galaxies at $z=1$ \citep{L11}.  At $z=0.5$, the typical covering fraction varies from $f_{c}=0.25$ to $f_{c}=1$, depending on the chosen limits of \brittion{Mg}{II} equivalent width and impact parameter \citep{BB91, Bechtold92, Steidel94, TB05}. In the local Universe, $f_{c}$ appears to drop below 0.25 for all galaxies \citep{BC09}, but remains high among those with high star-formation rates \citep{Bowen95}.  Thus, we seem to be observing a trend of a declining mean $f_{c}$ with time.  This is, of course, what one might expect given the fact that the global SFR density is also dropping dramatically from $z\sim2$.  As the fraction of galaxies with haloes rich in cold gas is depleted with time, the average $f_{c}$ should also be expected to decline.

If we include all galaxies detected within 150 kpc in this study, regardless of redshift or SFR, we estimate that the covering fraction for W$_{r}^{2796}>0.2$\AA~ is $f_{c}$(60 kpc) $=0.75$, $f_{c}$(100 kpc) $=0.66$, and $f_{c}$(150 kpc) $=0.63$.  

\section{Summary}

By exploiting the only known bright $z>2$ quasar within the 3D-HST Survey area, we have analyzed the W$_{r}^{\lambda2796}>0.2$\AA~ metal absorption properties of a small but complete sample of spectroscopically-confirmed galaxies with H$_{140}\gtrsim24$ at $0.65<z<2$.  
The unique depth and resolution of the WFC3 observations have enabled an examination of the host galaxies of quasar absorption lines over a range of impact parameters and SFRs previously unavailable at high redshift, and the slitless nature of the G141 grism ensures a nearly complete redshift census of galaxies proximate to the quasar sight line. Thus, our analysis avoids the most prohibitive selection biases of previous studies of absorber host galaxies at $z>1$.  

We unambiguously detect an isolated host galaxy for each of the three multi-ion absorption line systems detected with W$_{r}^{\lambda2796}>0.8$\AA~ along the line of sight to quasar SDSS J123622.93+621526.6.  The galaxies occupy the redshift range $1<z<2$ and have SFRs in excess of 4 M$_{\odot}$ yr$^{-1}$.  In each case the $\Sigma$SFR $>0.3$ M$_{\odot}$ yr$^{-1}$ kpc$^{-2}$, exceeding 0.1 M$_{\odot}$ yr$^{-1}$ kpc$^{-2}$, the threshold for local starburst galaxies. 

The three absorber host galaxies have a mean stellar mass of log(M$_{*}$/M$_{\odot})\sim9.75$, consistent with the typical dark matter halo masses derived from clustering measurements of \brittion{Mg}{II} absorbers at $z\lesssim1$.  The F140W morphologies of the host galaxies indicate a wide range in S{\'e}rsic indices, and profiles consistent with the compact triaxial systems and thick disks commonly observed among similarly low mass galaxies at $z\gtrsim2$.

We observe no strong evolution in the relation between \brittion{Mg}{II} equivalent width and impact parameter from $z\sim2$.  However, the spatial distribution of the high redshift \brittion{Mg}{II} absorption around the host galaxies appears to have a weaker azimuthal dependence than studies at $z\lesssim1$ \citep{Bordoloi11,Kacprzak11b,Kacprzak12}, indicating possible evolution in the angular distribution of high-W$_{r}$ \brittion{Mg}{II} absorption in galaxy haloes. Though a larger statistical sample would be required to confirm this behavior, these results are consistent with the recent findings of an increasing collimation of star formation-driven outflows with time from $z\sim3$, resulting from the evolution of low-mass star-forming galaxies from dispersion-dominated triaxial systems to stable disks \citep{Law12}. 

By examining the $W_{r}>0.2$\AA~ absorption properties of all galaxies detected in 3D-HST within 150 kpc of the quasar sight line we measure the typical covering fraction of \brittion{Mg}{II}-enriched gas at $z>0.65$ to be: f$_{c}=0.75$ at $\rho<60$ kpc, f$_{c}=0.66$ at $\rho<100$ kpc, and f$_{c}=0.63$ at $\rho<150$ kpc.  In addition, we find that the W$_{r}>0.8$\AA~ covering fraction is approximately unity to at least 60 kpc around star-forming galaxies at $1<z<2$.

Looking toward future work, we note that this analysis would greatly benefit from WFC3/IR follow-up of additional fields around absorber-rich sight lines.  The vast quasar catalogs available from the SDSS \citep{York2000, Schneider10} and the SDSS-III Baryon Oscillation Spectroscopic Survey \citep{Eisenstein11,Ross12,Paris12} are producing unprecedented numbers of metal absorption line systems \citep[e.g.,][]{York12}.  Targeting the densest of these sight lines with {\it HST} would likely provide the numbers of absorber-galaxy pairs to better constrain the evolution hinted at in this analysis.

\acknowledgments

This work would not have been possible without the publicly available data from the Sloan Digital Sky Survey, the Team Keck Redshift Survey  and from the GOODS-N grism program GO-11600 (PI Benjamin Weiner).   We are also grateful to the authors and maintainers of the AXE software package: Martin K$\ddot{u}$mmel, Harald Kuntschner, Jeremy Walsh, and Howard Bushouse.  We thank Glenn Kacprzak, Nikole Nielson, and Jason Tumlinson for helpful discussions. Support from HST grant GO-12177 is gratefully acknowledged.  We also acknowledge funding from ERC grant HIGHZ no. 227749.  This work is based on observations taken by the 3D-HST Treasury Program with the NASA/ESA HST, which is operated by the Association of Universities for Research in Astronomy, Inc., under NASA contract NAS5-26555. This research has made extensive use of NASAÕs Astrophysics Data System Bibliographic Services and of open source scientific Python libraries, including PyFITS and PyRAF produced by the Space Telescope Science Institute, which is operated by AURA for NASA.  



{\it Facilities:} \facility{HST (STIS)}, \facility{SSO (SSC)}.

\clearpage

\end{document}